\documentclass[11pt]{article}

\usepackage[margin=1in]{geometry}
\usepackage{amsmath,amssymb,amsfonts,amsthm}
\usepackage{bm}
\usepackage{enumerate}
\usepackage{float}
\usepackage[authoryear,round]{natbib}
\usepackage[hidelinks]{hyperref}

\theoremstyle{plain}
\newtheorem{theorem}{Theorem}
\newtheorem{proposition}{Proposition}
\newtheorem{lemma}{Lemma}

\theoremstyle{remark}

\theoremstyle{definition}
\newtheorem{definition}{Definition}

\title{Design-Driven Inference for Online Experiments:\\
Jointly Optimal Evidence Collection, Testing, and Estimation}
\author{Jun Yu\textsuperscript{1} \quad
Wenbiao Zhao\textsuperscript{2,*} \quad
Lixing Zhu\textsuperscript{3,*}\\[0.7em]
\small\textsuperscript{1}School of Mathematics and Statistics, Beijing Institute of Technology, Beijing 100081, China\\
\small\textsuperscript{2}School of Science, China University of Mining and Technology-Beijing, Beijing 100872, China\\
\small\textsuperscript{3}Department of Statistics, Beijing Normal University at Zhuhai, Zhuhai 519090, China\\[0.4em]
\small\textsuperscript{*}Corresponding authors: \href{mailto:wbzhao96@163.com}{wbzhao96@163.com}, \href{mailto:lzhu@bnu.edu.cn}{lzhu@bnu.edu.cn}}
\date{}

\begin{document}
\maketitle

\begin{abstract}
Online multifactorial experiments should allow early stopping for practically negligible effects while controlling false stopping under continuous monitoring and retaining precise treatment-effect estimation. We develop a design-driven framework that chooses allocation and the evidence rule jointly, treating allocation as evidence collection and testing and estimation as complementary uses of the same evidence. For multifactor experiments with nuisance block effects and treatment-by-block interactions, we show that block-orthogonal designs remove nuisance contamination from the treatment score and form a complete class for worst-case evidence growth. Under a fixed information budget, an isotropic allocation and an explicit radial e-value jointly attain the minimax rate optimal for directionally unknown alternatives. The same allocation is universally optimal for maximum likelihood estimation of treatment effects, simultaneously achieving A-, D-, and E-optimality and establishing double optimality for testing and estimation. Batchwise replication yields an anytime-valid e-process, retains minimax rate optimality for cumulative conditional worst-case evidence growth, and preserves universal estimation optimality for the prespecified complete experiment. Simulations and a large language model prompt experiment illustrate gains in stopping efficiency and estimation precision.
\end{abstract}

\section{Introduction}
\label{sec:intro}

Online experimentation has become a routine tool for evaluating product, advertising, and platform changes as user traffic arrives \citep{Larsen2024Statistical}. In a multifactor experiment with \(q\) treatment factors, the experimenter faces two coupled decisions: which treatment combinations receive scarce traffic, and when the accumulated evidence suffices to stop. The operational goal is asymmetric. If the treatment effect is too small to matter, the experiment should stop as soon as possible. If a meaningful effect exists, the experiment should continue, both to accumulate evidence and to estimate treatment effects precisely enough for downstream decisions. When the experiment does not stop early, the same procedure should also deliver efficient estimation, ideally satisfying standard optimality criteria such as A-, D-, and E-optimality. Stopping and estimation are therefore two arms of the same sequential design problem, not competing objectives. A satisfactory procedure must also remain valid under continuous monitoring: it must control the probability of incorrectly stopping a meaningful effect at arbitrary stopping times.

Constructing such a satisfactory procedure is structurally difficult. It appears on both the sequential
testing side and the design side. 
From the sequential testing side, the difficulty appears at two levels. The first is
nuisance. Multifactor experiments intrinsically contain block effects and
treatment-by-block interactions, which enter the same linear system as the
treatment effect. If the design is treated as exogenous, no post-design
method, including both \(p\)-process based or an e-value, can separate
treatment from nuisance once the allocation confounds them. Plugging in
nuisance estimates generally destroys validity, and under uncontrolled
allocation the evidence can lose first-order growth as the nuisance dimension
approaches the batch size. The second is the construction of an optimal
statistic. Classical \(t\)- and \(F\)-statistics are calibrated for a fixed
sample size and lose Type~I error control under continuous monitoring. The always-valid \(p\)-process of \citet{Johari2022always} restores calibration,
but calibration is not optimality. It remains unclear what the optimal
evidence rule is under continuous monitoring. 
Neither difficulty can be resolved on the
inference side alone, because both are created or constrained by the design.

On the design side, the difficulty arises when the inference procedure
is fixed in advance. No existing design theory delivers the stopping and
estimation goals jointly. This is because the allocation is usually forced to
serve a single objective, creating a conflict between parameter estimation
and testing: an allocation that is optimal for one need not be optimal for
the other.
Optimal design theory confirms this conflict. A-, D-, and E-optimality
target estimation while T- and KL-optimality target model discrimination. A design that is optimal for one
criterion need not be optimal for another, and this is particularly evident
when the two objectives are model discrimination and parameter estimation
\citep{Dette2005optimal,Ai2025JASA}. The root of this conflict lies in the
inference procedure itself: a testing procedure measures evidence through
power, whereas maximum likelihood measures evidence through parameter
location, and the two exploit the same data from different angles. As long
as the inference procedure is fixed in advance, the design can be aligned
with only one of them.

A distinctive feature of online experimentation is that the data scientist can actively choose both data collection and inference. Our approach follows from an evidence perspective, which we call \emph{design-driven inference}:
the design is chosen as part of the inferential procedure, not treated as an
exogenous input that inference must repair afterwards. At the design level, we treat allocation as evidence collection: by enforcing block orthogonality, treatment directions are made orthogonal to block effects and treatment-by-block interactions, so the evidence is collected free of nuisance contamination. From an inference perspective, we treat testing and estimation as two uses of the same evidence. Testing asks whether the evidence suffices to reject a negligible effect and estimation asks which effect values the evidence supports. We extract evidence through e-values and maximum likelihood, respectively. These two are natural choices from the evidence perspective: e-values describe how evidence accumulates against the null under continuous monitoring \citep{Shafer2021testing}, and the maximum likelihood estimator describes where the evidence locates the parameter \citep{Fisher1925Theory}. In this sense, the e-value plays a role analogous to that of the MLE: both are direct summaries of the evidence in the data, one from the testing angle and one from the estimation angle. 
Under this perspective, both objectives are expressions of the same underlying quantity, the evidence in the data, so a design that optimizes evidence collection can serve both. Moreover, because testing and estimation operate on the same evidence, evidence collection and evidence use can be optimized jointly rather than traded off.
Introducing e-values in our research is therefore not only a technical convenience but also places testing, estimation, and design in a single decision space.

\textbf{Related work}.
Always-valid sequential inference begins with Wald's sequential probability ratio test \citep{Wald1945seq}, which provides exact error control under continuous monitoring but assumes simple hypotheses, a single data stream, and no nuisance parameters. One line of subsequent work constructs confidence sequences and always-valid \(p\)-values, notably the \(p\)-processes of \citet{Johari2022always}, which are calibrated for rejection thresholds under continuous monitoring \citep{Howard2021uniform}. A parallel line instead builds evidence from e-values and e-processes, which compose multiplicatively and admit an expected log-growth criterion. Within this e-value framework, nuisance is handled either by Bayesian mixtures that integrate over a prior \citep{grunwald2020safe,vovk2020e-value} or by minimax and least-favourable constructions that avoid a subjective prior \citep{Shafer2021testing,perezortiz2024estatistics}. Both lines, however, treat the allocation as fixed: they preserve validity for a given design rather than optimize the design jointly with the evidence rule.

Classical optimal design theory separates into estimation and discrimination. A-, D-, and E-optimality target estimation precision \citep{atkinson2007sas} while T- and KL-optimality \citep{atkinson1975design,LopezFidalgo2007KL,dette2009optimal}, target model discrimination, and DT, DKL-optimality compromise between discrimination and estimation \citep{ATKINSON2008DT,May2013DKL}. Both strands assume a fixed inferential procedure and are generally mutually inconsistent: a design that is optimal for one criterion need not be optimal for another. Our formulation instead chooses allocation and e-value jointly, so that evidence growth and terminal estimation are aligned rather than traded off.

\textbf{Our contributions}.
We make three main contributions, all from an evidence perspective: design
collects evidence, testing and estimation use it, and e-values and maximum
likelihood extract it.

Our first contribution identifies the structural obstruction to anytime-valid
inference in multifactor experiments and shows that it can be removed by
design. If the treatment and nuisance column spaces overlap, the worst-case
expected log-growth is zero. Even when identifiability holds, nuisance
variation can reduce the attainable growth below the isotropic benchmark. We then show that block-orthogonal designs, within the reassignment-feasible orbit natural in online experiments, form a complete class without loss of optimality, and derive the optimal allocation and e-value jointly in explicit closed form (Theorem~\ref{thm:exact-complete-class}).

Our second contribution is a double-optimality principle that unifies testing
and estimation. The same isotropic design that is optimal for evidence
accumulation is also universally optimal for estimation, simultaneously
achieving A-, D-, and E-optimality, and this design--e-value pair remains
jointly first-order minimax-rate optimal when competing designs are allowed
to be arbitrary, subject only to estimability and a natural information
budget (Theorem~\ref{thm:minimax-opt}). The design-level tension between
testing and estimation therefore disappears once both are viewed as uses of
the same evidence.

Our third contribution extends this double optimality to the sequential
setting through batchwise additivity. The resulting replicative construction
is simultaneously anytime-valid, asymptotically optimal, and
estimation-optimal, and adaptive reallocation cannot improve the
least-favourable direction-free local information
(Theorem~\ref{thm:sequential-grow}).

Methodologically, the paper replaces the conventional two-step
practice, which fixes a design, then constructs a test, by a single decision problem
in which allocation and evidence rule are chosen jointly. This yields a
precomputable, replicable schedule that is simultaneously anytime-valid and
estimation-optimal. Numerical studies in Sections~\ref{sec:numerical} and
\ref{sec:real-data} evaluate the proposed procedures against standard
benchmark designs, examining empirical Type~I error under continuous
monitoring, expected stopping time, and post-experiment estimation accuracy.
A real-data application illustrates the practical gains of the joint
design--inference framework.

\vspace{-1em}
\section{Preliminaries}\label{sec:model}
\subsection{Problem Setup}\label{subsec:model}
Consider a multifactor online experiment with controllable treatment allocation. To avoid confounding the treatment effects of interest with known sources of heterogeneity, such as time, user cohort, or platform, we partition the experimental factors into two groups. The treatment factors of interest, \(\mathbf{x}\in\mathbb{R}^q\),  are the attributes the data scientist wishes to evaluate; 
the blocking factors, \(\mathbf{z}\in\mathbb{R}^{p-q}\), capture remaining systematic variation. 
{In this paper, \(q\) is assumed to be fixed while \(p\) is allowed to diverge with respect to the sample size of each batch. This reflects the practical situation where the number of experimental units in each batch is typically not much larger than the total number of factors under study.}
Across \(n\) runs, their rows form the design matrices \(\mathbf X\) and \(\mathbf Z\), while \(\mathbf J\) collects the treatment-by-block interaction columns. The choice of these matrices is part of the inferential problem rather than an assumption imposed at this stage.

The observed response follows the linear model \vspace{-1em}
\begin{equation}\label{eq:model}
Y = \mathbf{x}^\top \boldsymbol\psi + \mathbf{z}^\top \boldsymbol\lambda + (\mathbf{x}\otimes\mathbf{z})^\top \boldsymbol\theta + \varepsilon,\vspace{-1em}
\end{equation}
with \(\varepsilon\sim N(0,1)\), where \(\boldsymbol\psi\in\mathbb{R}^q\) is the treatment effect vector of interest, \(\boldsymbol\lambda\in\mathbb{R}^{p-q}\) collects the blocking (nuisance) main effects, and \(\boldsymbol\theta\) captures treatment-by-block interactions, also treated as nuisance. The unit variance assumption entails no loss of generality, as the variance of the error terms can be absorbed by scaling of the design matrix.

The testing problem is therefore \vspace{-1em}
\begin{equation}\label{eq:hypothesis}
\mathcal{H}_0:\|\boldsymbol\psi\|\ge\Delta
\quad\text{against}\quad
\mathcal{H}_1:\|\boldsymbol\psi\|<\Delta,\vspace{-1em}
\end{equation}
where \(\Delta>0\) is a minimum effect size of practical interest, prespecified by the manager. Stopping the experiment means concluding that the treatment effect lies inside the practical-equivalence ball; Type-I error is the probability of making that conclusion when the effect norm is at least \(\Delta\).

The procedure must meet two objectives that are naturally in tension. \\
\textbf{Objective I}: under continuous monitoring, it must provide anytime-valid Type-I error control.
That is to say, the probability of ever rejecting \(\mathcal{H}_0\) under the null must remain bounded by \(\alpha\), regardless of the stopping rule employed. \\
\textbf{Objective II}: if the experiment reaches its planned horizon without early termination, the collected data must support precise estimation of \(\boldsymbol\psi\) for downstream operational decisions. \\Reconciling these two objectives: fast sequential testing with error guarantees and clean post-experiment estimation, is the central challenge our design-driven framework addresses.


\subsection{The e-Value and GRO Criterion}
\label{subsec:evalue}

An {e-variable} is a nonnegative random variable \(E\) satisfying \vspace{-1em}
\begin{equation}
    \mathbb{E}_{P}[E] \le 1 \quad \text{for all } P \in \mathcal{H}_0,\vspace{-1em}
\end{equation}
with its realized value called an {e-value}. This simple expectation constraint has a direct consequence for Type-I error control: by Markov's inequality, rejecting \(\mathcal{H}_0\) when \(E \ge 1/\alpha\) yields a test with error probability bounded by \(\alpha\). The reciprocal \(1/E\) is a conservative \(p\)-value \citep{vovk2020e-value}.

An e‑value is interpreted as a bet against the null. 
Let \(E_{(j)}\) denotes the conditional \(e\)-variable for the \(j\)th batch of data, satisfying \(\mathbb{E}(E_{(j)} \mid \mathcal{F}_{j-1}) \le 1\) under the null. 
Their product \(E^{(m)} = \prod_{j=1}^m E_{(j)}\) defines the cumulative evidence process (a nonnegative test supermartingale).
Under optional continuation, the evidence accumulates multiplicatively, and its long‑run behaviour is governed by the asymptotic logarithmic growth rate: under mild conditions, the law of large numbers yields  
\(
{m}^{-1}\log E^{(m)} \;\to\; \mathbb{E}[\log E] \quad (m\to\infty),
\)  
so that maximizing the expected logarithm is equivalent to maximizing the exponential growth rate of the capital per study, i.e. the doubling rate of Kelly betting \citep{kelly1956,Shafer2021testing}.  
This time‑average optimality, rather than a static single‑study criterion, motivates the term growth‑rate optimality, which is adopted as a measure of the power for e-values-based test problems.

For a composite alternative \(\mathcal{H}_1 = \{P_\theta : \theta \in \Theta_1\}\), \cite{grunwald2020safe} define the  growth‑rate optimal in the worst case (GROW): \vspace{-1em} 
\[
\sup_{E\in\mathcal{E}(\Theta_0)}\inf_{\theta\in\Theta_1} \mathbb{E}_{P_\theta}[\log E], \vspace{-1em} 
\]  
where \(\mathcal{E}(\Theta_0)\) denotes the class of all e‑variables for the null. In the remainder of this paper, we opt to use $\mathbb{E}_{\theta}$ instead of $\mathbb{E}_{P_\theta}$ to emphasize the dependence of parameter.
In the testing problem  \eqref{eq:hypothesis}, the alternative \(\mathcal{H}_1=\{\boldsymbol\psi:\|\boldsymbol\psi\|<\Delta\}\) contains points such that \(\|\boldsymbol\psi\|\) is arbitrarily close to $\Delta$. 
At these points the null-boundary condition forces any e‑variable to satisfy \(\mathbb{E}_{P_{\mathbf{0}}}[\log E]\le 0\); hence the trivial e‑variable \(E\equiv 1\) attains the worst‑case expected log‑growth \(0\). 
More mathematical details can be found in Proposition~\ref{prop:fail}.
To obtain a meaningful optimality benchmark near the boundary we introduce a sequence of separated alternatives \vspace{-1em} 
\[
\mathcal{H}_{1,n}(\rho_n) = \{\boldsymbol\psi: \|\boldsymbol\psi\| \le \Delta - \rho_n\}, \qquad 0 \le \rho_n \le \Delta, \vspace{-1em} 
\]
where the separation $\rho_n$ may depend on the sample size $n$.  Let $\Theta_{1,n}(\rho_n)$ be the full parameter space under the alternative, i.e., the set of all $(\boldsymbol\psi,\boldsymbol\lambda,\boldsymbol\theta)$ with $\|\boldsymbol\psi\| \le \Delta - \rho_n$.  For fixed $n$ and $\rho_n$, an e‑variable $E$ is called {local GROW} if it attains \vspace{-1em} 
\begin{equation}\label{eq:localGROW}
\sup_{E \in \mathcal{E}_n(\mathcal{H}_0)} \inf_{(\boldsymbol\psi,\boldsymbol\lambda,\boldsymbol\theta) \in \Theta_{1,n}(\rho_n)} \mathbb{E}_{\boldsymbol\psi,\boldsymbol\lambda,\boldsymbol\theta}[\log E], \vspace{-1em} 
\end{equation}
where $\mathcal{E}_n(\mathcal{H}_0)$ denotes the class of e‑variables for $\mathcal{H}_0$ that are measurable with respect to the first $n$ observations.  Because the alternative boundary is now strictly separated from the null, the trivial e‑variable is sub‑optimal, and the maximin problem yields a non‑trivial benchmark.  As $\rho_n \to 0$ the optimal expected log‑growth tends to zero, and its rate of decay with $n$ characterizes the efficiency limit of evidence accumulation near the boundary of the practically‑equivalent region. Therefore, the remainder of our study is conducted entirely within the local GROW framework.

\section{Necessary Conditions of Design and Optimal e-value}\label{subsec:e-value-comp}

We work in a one‑period setup with a fixed sample size $n$ and a controllable design.  
To construct a non‑trivial e‑variable with power under~\eqref{eq:hypothesis}, two obstacles must be faced.  
First, plug‑in estimation of the nuisance parameters $(\boldsymbol\lambda,\boldsymbol\theta)$ typically violates the e‑value property, because the resulting statistic need not have expectation at most one uniformly over $\mathcal{H}_0$.  
Second, the standard recipe for growth‑rate optimality, which relies on a least favourable likelihood ratio between the null and alternative classes via reverse information projection \citep{grunwald2020safe}, degenerates to the trivial e‑variable $E\equiv1$ when unrestricted nuisance parameters are present (see Proposition~\ref{prop:fail} for details).
Under uncontrolled i.i.d.\ sampling, these difficulties are hard to solve, but when the treatment allocation is under the experimenter's control, algebraic structure can be imposed to decouple the interest parameter from the nuisance components.

\subsection{Necessary Conditions of Design}\label{subsec:necessary-design}

Write \(\mathbf W=[\mathbf Z,\mathbf J]\), with \(\mathbf J\) denoting the
treatment-by-block interaction terms, for the combined nuisance design
matrix. The treatment effect is not identifiable from the full data if a
nonzero treatment direction can be represented exactly by a nuisance
direction. The following theorem formalizes the resulting necessary design
condition.

\begin{theorem}[Necessary design condition]
\label{thm:identifiability}
Assume that \(\mathbf X\) has full column rank. Consider the condition
\begin{equation}\label{eq:identifiability-condition}
\operatorname{col}(\mathbf X)
\cap
\operatorname{col}([\mathbf Z,\mathbf J])
=\{\mathbf0\}.
\end{equation}
If \eqref{eq:identifiability-condition} fails, then for every
\(0<\rho_n\le\Delta\) there exist a null parameter
\((\boldsymbol\psi_0,\boldsymbol\lambda_0,\boldsymbol\theta_0)\) and an
alternative parameter
\((\boldsymbol\psi_1,\boldsymbol\lambda_1,\boldsymbol\theta_1)\in
\Theta_{1,n}(\rho_n)\) that induce the same full-data distribution.
Consequently,
\begin{equation}\label{eq:full-data-grow-zero}
\sup_{E\in\mathcal E_n(\mathcal H_0)}
\inf_{(\boldsymbol\psi,\boldsymbol\lambda,\boldsymbol\theta)
\in\Theta_{1,n}(\rho_n)}
\mathbb E_{\boldsymbol\psi,\boldsymbol\lambda,\boldsymbol\theta}
[\log E]
=0.
\end{equation}
\end{theorem}

Theorem~\ref{thm:identifiability} identifies a structural obstruction: if
\(\operatorname{col}(\mathbf{X}) \cap \operatorname{col}([\mathbf{Z},\mathbf{J}]) \neq \{\mathbf{0}\}\),
there exist indistinguishable null and alternative configurations, forcing
every e-variable to have non-positive worst-case expected log-growth.
A sufficient way to remove this intersection is to make the column spaces
orthogonal, i.e.\ to impose block orthogonality
\(\mathbf{X}^\top\mathbf{Z}=\mathbf{0}\) and
\(\mathbf{X}^\top\mathbf{J}=\mathbf{0}\).
In online experiments, different treatment configurations are typically run
concurrently within the same time period, so that traffic-splitting platforms
naturally yield balanced sample sizes across configurations. We compare
allocations that preserve both the treatment rows and the blocking rows, but
allow the blocking rows to be reassigned to treatment slots. This corresponds,
for example, to assigning treatments within a prespecified collection of
blocks.

However, identifiability alone does not guarantee first-order evidence growth:
nuisance variation can still degrade the attainable worst-case log-growth even
when the treatment and nuisance spaces are distinguishable. The following
proposition makes this point precise in an uncontrolled i.i.d.\ setting.
\vspace{-2em}
\begin{proposition}\label{prop:fail}
Under model~\eqref{eq:model}, consider uncontrolled i.i.d.\ sampling without
design-imposed orthogonality. Let \(q\) be fixed and \((p-q)/n\to1\).
The rows \((\mathbf x_i,\mathbf z_i)\) of \((\mathbf X,\mathbf Z)\) are
i.i.d., with \(\mathbf x_i\in[-1,1]^q\) mean zero and independent of
\(\mathbf z_i\), whose coordinates are i.i.d.\ from a nondegenerate
absolutely continuous distribution on \([-1,1]\). Then, for every
\(0<\rho_n\le\Delta\), the conditional version of the local GROW criterion
defined in Equation~\eqref{eq:localGROW} satisfies \vspace{-1em}
\[
\sup_{E\in\mathcal E_n(\mathcal H_0)}
\inf_{(\boldsymbol\psi,\boldsymbol\lambda,\boldsymbol\theta)
\in\Theta_{1,n}(\rho_n)}
\mathbb E_{\boldsymbol\psi,\boldsymbol\lambda,\boldsymbol\theta}
\!\left[\log E\,\middle|\,\mathbf X,\mathbf Z\right]
=o_p(n\rho_n^2).
\]
\end{proposition}

Proposition~\ref{prop:fail} shows that even when identifiability holds, the
conditional local GROW optimum can be \(o_p(n\rho_n^2)\), below the test power of the proposed method (will be shown in Section~\ref{sec:joint}) in the sense of local power rate. Thus block orthogonality serves a second role
beyond identifiability: it isolates nuisance from the treatment score and
restores the first-order rate. We now formalize the class of designs that
achieve this isolation by reassignment.

We note that the information projection construction of
\cite{grunwald2020safe}, which underlies the REGROW e-variable, can be
specialised to our local alternative setup with the penalty function
\(f\equiv 0\) and applied to the conditional model given the realised design
\([\mathbf X,\mathbf Z]\). In that case, it targets the conditional local
GROW optimum \vspace{-1em}
\[
\sup_{E\in\mathcal E_n(\mathcal H_0)}
\inf_{(\boldsymbol\psi,\boldsymbol\lambda,\boldsymbol\theta)\in\Theta_{1,n}(\rho_n)}
\mathbb E_{\boldsymbol\psi,\boldsymbol\lambda,\boldsymbol\theta}
\!\left[\log E\,\middle|\,\mathbf X,\mathbf Z\right].\vspace{-1em}
\]
Proposition~\ref{prop:fail} shows that even the oracle value of this
conditional local GROW criterion is \(o_p(n\rho_n^2)\), and hence lies below
the minimax benchmark \(B_n(\rho_n)\). Thus, although the reverse information
projection method is designed to attain the conditional local GROW optimum,
it may fail to be minimax rate-optimal under uncontrolled random designs
without design-imposed orthogonality. Besides this, information projection
over the composite hypotheses and unrestricted nuisance parameters is
high-dimensional and generally does not yield a closed-form e-value, while our
design-driven method directly yields an explicit solution, which makes the
computation much faster.

Precisely, let \(D=(\mathbf X,\mathbf Z,\mathbf J)\), let
\(\mathcal X(D)\) be the set of distinct treatment rows, and write
\(I_{\mathbf x}=\{i:\mathbf x_i=\mathbf x\}\). For a permutation
\(\pi\) of \(\{1,\ldots,n\}\), let \(\mathbf Z^\pi\) have row
\(\mathbf z_{\pi(i)}\) in position \(i\). Define
\(\mathfrak D_n^{\mathrm{balance}}\) as the class of designs for which
there exists such a permutation satisfying
\begin{equation}\label{eq:conditional-block-balance}
\sum_{i\in I_{\mathbf x}}\mathbf z_{\pi(i)} = \mathbf 0
\quad\text{for every }\mathbf x\in\mathcal X(D).
\end{equation}
For \(D\in\mathfrak D_n^{\mathrm{balance}}\), fix one such permutation,
put \(\mathbf Z^\perp=\mathbf Z^\pi\), retain the treatment matrix
\(\mathbf X\), and define row \(i\) of \(\mathbf J^\perp\) as
\(\mathbf x_i\otimes \mathbf z_i^\perp\). Thus
\(D^\perp=(\mathbf X,\mathbf Z^\perp,\mathbf J^\perp)\) preserves the
two marginal row multisets and changes only their pairing.

A concrete sufficient condition is available for two-level blocking
factors. Put \(d=p-q\). Suppose every
\(\mathbf z\in\{-1,1\}^d\) occurs exactly \(r\) times in the blocking
multiset and every distinct treatment row \(\mathbf x\) occurs
\(m_{\mathbf x}2^d\) times, where \(r\) and all \(m_{\mathbf x}\) are
nonnegative integers satisfying
\(\sum_{\mathbf x}m_{\mathbf x}=r\). Assigning
\(m_{\mathbf x}\) copies of every blocking profile to treatment row
\(\mathbf x\) then satisfies~\eqref{eq:conditional-block-balance}.

For a fixed design \(D\), let \(\mathcal E_n(\mathcal H_0;D)\) denote
the full-data e-variables valid under \(D\), and define
\[
\gamma_n(D,E;\rho_n)
=
\inf_{(\boldsymbol\psi,\boldsymbol\lambda,\boldsymbol\theta)
\in\Theta_{1,n}(\rho_n)}
\mathbb E^D_{\boldsymbol\psi,\boldsymbol\lambda,\boldsymbol\theta}
[\log E].
\]
Recall that block orthogonality discussed below
Theorem~\ref{thm:identifiability} is not the only way to satisfy the
necessary condition, but it is the one that is compatible with the
reassignment operation natural in online experiments. The following theorem
shows that it in fact exhausts the optimal designs within that orbit.

\begin{theorem}[Complete class]
\label{thm:exact-complete-class}
Assume model~\eqref{eq:model} with unrestricted nuisance parameters and
that \(\mathbf X\) has full column rank. For every
\(D\in\mathfrak D_n^{\mathrm{balance}}\) and every
\(E\in\mathcal E_n(\mathcal H_0;D)\), there exist the block-orthogonal
reassignment \(D^\perp\in\mathfrak D_n^{\mathrm{balance}}\) described
above and a score-based e-variable
\(E^\perp\in\mathcal E_n(\mathcal H_0;D^\perp)\) such that \vspace{-1em}
\begin{equation}\label{eq:exact-complete-dominance}
\gamma_n(D,E;\rho_n)
\le
\gamma_n(D^\perp,E^\perp;\rho_n)
\qquad(0<\rho_n\le\Delta).\vspace{-1em}
\end{equation}
Consequently, within every reassignment-feasible design orbit,
block-orthogonal designs form a complete class for worst-case expected
log-growth.
\end{theorem}

Within every reassignment-feasible design orbit, block-orthogonal designs
form a complete class for worst-case expected log-growth. In other words,
the nuisance isolation achieved by block orthogonality is not a restrictive
modelling choice: within the natural reassignment operation of online
experiments, it is without loss of optimality.

Block orthogonality enables a direct design--inference synergy. First, it
makes the treatment score \(\mathbf{S}_n=\mathbf{X}^\top\mathbf{Y}\)
sufficient for \(\boldsymbol\psi\) with a nuisance-free distribution,
reducing the e-value problem to the low-dimensional score. Second, it
simplifies the construction of the optimal e-value, yielding an explicit
closed form (introduced in Section~\ref{subsec:opt-e-con}).
Section~\ref{sec:joint} further proves that this e-value is minimax optimal,
a property that holds even for designs outside \(\mathfrak D_n^{\mathrm{balance}}\).
Thus block orthogonality simultaneously eliminates nuisance confounding and
supports a closed-form optimal safe test, making the design an active
component of inference rather than a mere precaution.

\subsection{Joint optimal design and inference}\label{subsec:opt-e-con}

We now determine the design of $\mathbf{X}$ that maximizes the local GROW criterion \eqref{eq:localGROW} among all block‑orthogonal allocations.  
Recall that, under block orthogonality, $\mathbf{S}_n \sim N(\mathbf{M}_n\boldsymbol\psi,\;\mathbf{M}_n)$ with $\mathbf{M}_n = \mathbf{X}^\top\mathbf{X}\succ \mathbf{0}$.  
The magnitude of $\mathbf{M}_n$ grows with $n$; we therefore write $\mathbf{M}_n = n\mathbf{A}_n$ where $\mathbf{A}_n\succ\mathbf{0}$ and, to fix the scale, $\operatorname{tr}(\mathbf{A}_n)=q$.  
The matrix $\mathbf{A}_n$ captures the allocation of information across directions of $\boldsymbol\psi$.  
Define the optimal worst‑case expected log‑growth for a given design $\mathbf{A}_n$ by \vspace{-1em}
\begin{equation}\label{eq:Gamma-def}
\Gamma_n(\mathbf{A}_n,\rho_n)
= \sup_{E\in\mathcal{E}_n(\mathcal{H}_0;\mathbf{A}_n)}
\inf_{\|\boldsymbol\psi\|\le\Delta-\rho_n}
\mathbb{E}_{\boldsymbol\psi}[\log E],\vspace{-1em}
\end{equation}
where $\mathcal{E}_n(\mathcal{H}_0;\mathbf{A}_n)$ is the class of e‑variables for $\mathcal{H}_0$ that are measurable with respect to the first $n$ observations under the design $\mathbf{A}_n$.

The following theorem shows that the isotropic design with $\mathbf{A}_n=\mathbf{I}_q$ is optimal, which guides us to find the design on $\mathbf{X}$. 

\vspace{-1em}

\begin{theorem}[Isotropic designs are asymptotically optimal]\label{thm:radial-reduction}
For every block‑orthogonal design satisfying the trace constraint and every $0\le\rho_n\le\Delta$, \vspace{-1em} 
\begin{equation}\label{eq:general-design-upper}
\Gamma_n(\mathbf{A}_n,\rho_n)
\le \frac{n\rho_n^2}{2}\lambda_{\min}(\mathbf{A}_n)
\le \frac{n\rho_n^2}{2}.\vspace{-1em}
\end{equation}
For every positive sequence $\rho_n\to0$,\vspace{-1em}
\begin{equation}\label{eq:isotropic-complete}
\frac{\sup_{\mathbf{A}_n\succ\mathbf{0}:\operatorname{tr}(\mathbf{A}_n)=q} \Gamma_n(\mathbf{A}_n,\rho_n)}{\Gamma_n(\mathbf{I}_q,\rho_n)} \longrightarrow 1.
\end{equation}

\end{theorem}

The information bound has a clear geometric interpretation. For a unit direction $\mathbf{u}$, the boundary pair $\boldsymbol\psi_0=\Delta\mathbf{u}$, $\boldsymbol\psi_1=(\Delta-\rho_n)\mathbf{u}$ gives score‑space Kullback–Leibler divergence $\frac{n\rho_n^2}{2}\mathbf{u}^\top\mathbf{A}_n\mathbf{u}$. Taking the least favourable direction yields the first inequality, and the trace constraint forces $\lambda_{\min}(\mathbf{A}_n)\le 1$, with equality only for the isotropic design. Thus isotropy is the unique way to avoid a weak local direction: any other allocation concentrates information in some directions at the expense of others, reducing the worst‑case growth rate.  

Theorem~\ref{thm:radial-reduction} shows that the isotropic design is asymptotically complete: as the separation shrinks, the ratio of the optimal growth rate under an arbitrary design to that under the isotropic design tends to one. Consequently, one can restrict attention to isotropic designs without first‑order asymptotic loss.


Motivated by the identifiability requirement in Theorem~\ref{thm:identifiability} and the design bound in Theorem~\ref{thm:radial-reduction}, we now construct the optimal e-value under the block-orthogonal isotropic design.
To be precise, under $\mathbf{A}_n = \mathbf{I}_q$, the score is $\mathbf{S}_n \sim N(n\boldsymbol\psi,\, n\mathbf{I}_q)$.  
The two hypotheses are invariant under rotations of $\boldsymbol\psi$, so a natural e‑variable is the spherical mixture likelihood ratio.  
Let $f_{q,\delta}$ denote the non‑central chi‑square density with $q$ degrees of freedom and non‑centrality $\delta$, evaluated at $Q_n = \|\mathbf{S}_n\|^2/n$.  
Define
\[
E^*_n = \frac{\int_{\|\boldsymbol\psi\|=\Delta-\rho_n} \exp\{-\frac{n}{2}\|\mathbf{S}_n/n-\boldsymbol\psi\|^2\}\,\mathrm{d}\sigma(\boldsymbol\psi)}
{\int_{\|\boldsymbol\psi\|=\Delta} \exp\{-\frac{n}{2}\|\mathbf{S}_n/n-\boldsymbol\psi\|^2\}\,\mathrm{d}\sigma(\boldsymbol\psi)}
= \frac{f_{q,\,n(\Delta-\rho_n)^2}(Q_n)}{f_{q,\,n\Delta^2}(Q_n)},
\]
where $\mathrm{d}\sigma$ is the uniform probability measure on the sphere.  
The following theorem states that $E^*_n$ is an optimal e-value when we adopt the design satisfying \(\mathbf{X}^\top\mathbf{Z}=\mathbf{0}\), \(\mathbf{X}^\top\mathbf{J}=\mathbf{0}\), and $\mathbf{M}_n=n\mathbf{I}_q$.

\begin{theorem}[Optimal e‑value under isotropic design]\label{thm:optimal-e}
Assume the design satisfying \(\mathbf{X}^\top\mathbf{Z}=\mathbf{0}\), \(\mathbf{X}^\top\mathbf{J}=\mathbf{0}\), and $\mathbf{M}_n=n\mathbf{I}_q$.  Then  
\begin{enumerate}
    \item ~~the e-value can be recast to \vspace{-1em}
    \begin{equation}\label{eq:bessel-opt-e}
E_n^* = 
\left(\frac{\Delta}{\Delta-\rho_n}\right)^{\!\frac{q-2}{2}}
\exp\!\Big(n\Delta\rho_n - \frac{n\rho_n^2}{2}\Big)
\,
\frac{I_{q/2-1}\!\big((\Delta-\rho_n)\sqrt{nQ_n}\big)}
     {I_{q/2-1}\!\big(\Delta\sqrt{nQ_n}\big)}, \vspace{-1em}
    \end{equation}
    with \(I_\nu\) being the modified Bessel function of the first kind. \vspace{-1em}
\item ~~the $E_n^*$ is an e‑value and is locally GROW optimal. 
\end{enumerate}
\end{theorem}

\section{Double Optimality for Testing and Estimation}
\label{sec:joint}

The design principles of Section~\ref{subsec:e-value-comp} narrow the allocation to block-orthogonal isotropic designs and yield an explicit e‑value.  
This raises two questions that we address in the present section.  
First, is the e‑value in Theorem~\ref{thm:optimal-e} optimal for accumulating evidence, thereby fulfilling {Objective~I} of Section~{subsec:model}. 
Second, does the same design permit optimal estimation of \(\boldsymbol\psi\), so that repeating it provides a principled route to {Objective~II}.  
We answer both affirmatively: the e‑value is locally minimax in terms of GROW, and the least‑squares estimator \(\hat{\boldsymbol\psi}=\mathbf{S}_n/n\) attains universal optimality \citep{kiefer1974general}, establishing a double optimality that unifies safe testing and efficient estimation under the proposed design and inference procedure.

\subsection{Minimax Optimality for Testing}\label{sec:detection-boundary}

In the previous section, under block orthogonality, we derived an
approximately optimal design and the optimal e-value. These two objects
jointly determine the local GROW rate, and the sharp attainable benchmark
is the radial quantity
\[
B_n(\rho_n)
:=
\mathrm{KL}\!\left(
\chi_q^2\!\left(n(\Delta-\rho_n)^2\right)
\,\middle\|\,
\chi_q^2\!\left(n\Delta^2\right)
\right),
\]
which satisfies
\[
B_n(\rho_n)
=
\frac{n\rho_n^2}{2}
\Big\{1+O(\rho_n)+O(n^{-1})\Big\}.
\]

A natural question now arises. The block-orthogonal construction was
derived under the constraint \(\mathbf X^\top\mathbf Z=\mathbf0\) and
\(\mathbf X^\top\mathbf J=\mathbf0\). Is this design--inference pair still
optimal when the constraint is removed? To answer this question, we first
recall a basic information-theoretic upper bound that holds for any
possible e-variable.

\begin{lemma}[Entropy bound for e-values]\label{lem:kl-bound}
Let \(\operatorname{conv}(\mathcal{H}_0)\) be the convex hull of the set of null distributions.
If \(E\) is valid for \(\mathcal{H}_0\), \(Q\) is any probability distribution and \(P\in\operatorname{conv}(\mathcal{H}_0)\), then
\(
\mathbb{E}_Q[\log E] \le {\rm KL}(Q\|P).
\)
Consequently,
\[
\inf_{Q\in\mathcal{H}_1}\mathbb{E}_Q[\log E]
\le
\inf_{Q\in\mathcal{H}_1}\inf_{P\in\operatorname{conv}(\mathcal{H}_0)} {\rm KL}(Q\|P).
\]
\end{lemma}

Under a block-orthogonal design, the Gaussian orthogonal decomposition
used in the proof of Theorem~\ref{thm:optimal-e} shows that any e-variable
can be replaced by one depending only on \(\mathbf{S}_n\) without reducing
the worst-case expected log-growth; it therefore suffices to consider
score-based e-variables.
For any such \(E\) and any alternative \(\boldsymbol\psi_1\) with
\(\|\boldsymbol\psi_1\|\le\Delta-\rho_n\), Lemma~\ref{lem:kl-bound} gives
\[
\mathbb{E}_{\boldsymbol\psi_1}[\log E]
\;\le\;
\inf_{\|\boldsymbol\psi_0\|\ge\Delta}
\mathrm{KL}\bigl(P_{\boldsymbol\psi_1}\,\|\,P_{\boldsymbol\psi_0}\bigr),
\]
where \(P_{\boldsymbol\psi}\) denotes the score distribution
\(N(n\mathbf{A}_n\boldsymbol\psi,n\mathbf{A}_n)\).
For a unit vector \(\mathbf{u}\), take the boundary pair
\(\boldsymbol\psi_0=\Delta\mathbf{u}\) and
\(\boldsymbol\psi_1=(\Delta-\rho_n)\mathbf{u}\); their
Kullback--Leibler divergence equals
\(\frac{n\rho_n^2}{2}\mathbf{u}^\top\mathbf{A}_n\mathbf{u}\).
Minimizing over \(\mathbf{u}\) and using
\(\operatorname{tr}(\mathbf{A}_n)=q\), which implies
\(\lambda_{\min}(\mathbf{A}_n)\le 1\), we obtain, for every
block-orthogonal design,
\[
\inf_{\|\boldsymbol\psi\|\le\Delta-\rho_n} \mathbb{E}_{\boldsymbol\psi}[\log E]
\;\le\; \frac{n\rho_n^2}2.
\]
Under isotropy, the sharper, attainable upper bound is the radial benchmark
\(B_n(\rho_n)\) defined above.

Thus, within the block-orthogonal isotropic class, the Bessel-ratio
e-value \(E_n^*\) defined in Theorem~\ref{thm:optimal-e} attains the
radial benchmark \(B_n(\rho_n)\). It remains to ask whether this
design--inference pair is optimal when the block-orthogonality constraint
is removed and competing designs are allowed to be arbitrary.

To formalize this, let
\(\widetilde D=(\widetilde{\mathbf X},\widetilde{\mathbf Z},
\widetilde{\mathbf J})\) range over all competing designs without assuming
block orthogonality. Assume that \(\widetilde D\) satisfies
\(\widetilde{\mathbf X}^{\top}\widetilde{\mathbf X}
=n\widetilde{\mathbf A}_n\).
The condition \(\widetilde{\mathbf A}_n\succ\mathbf0\) ensures that
\(\boldsymbol\psi\) is estimable, while
\(\operatorname{tr}(\widetilde{\mathbf A}_n)\le q\) is a natural resource
constraint: in a standard \(\pm1\) factorial design each row satisfies
\(\mathbf{x}_i^\top\mathbf{x}_i=q\), so this constraint fixes the total
information budget and ensures a fair comparison across designs. Let
\(D=(\mathbf X,\mathbf Z,\mathbf J)\) be a block-orthogonal isotropic
design satisfying \(\mathbf X^\top\mathbf Z=\mathbf0\),
\(\mathbf X^\top\mathbf J=\mathbf0\), and
\(\mathbf X^\top\mathbf X=n\mathbf I_q\), and let \(E_n^*\) be the
Bessel-ratio e-value defined in Theorem~\ref{thm:optimal-e}.

\begin{theorem}[Minimax rate optimality]\label{thm:minimax-opt}
If $\rho_n>0$ and
$\rho_n\to0$, then we have
\[
\frac{\gamma_n(D,E_n^*;\rho_n)}
{\displaystyle\sup_{\widetilde D}
\ \sup_{E\in\mathcal E_n(\mathcal H_0;\widetilde D)}
\gamma_n(\widetilde D,E;\rho_n)}
\longrightarrow1.
\]
Thus $D$ and $E_n^*$ are jointly minimax-rate optimal over the
stated design class and all corresponding full-data e-values.
\end{theorem}
\vspace{-1em}
When \(\rho_n \to \infty\) the separation between the null and alternative
spheres is large enough for the evidence to diverge and consistent
detection to be possible.
In the boundary-local regime \(\rho_n \to h>0\) the radial statistic
converges to a Gaussian shift.
If \(\rho_n\to0\), then \(B_n(\rho_n)\to0\) and no e-value can guarantee a
positive worst-case growth. In this regime the null and alternative
boundaries merge, while the exact Bessel-ratio e-value tends to one in
probability under both boundary laws. Thus, we do not take this scenario
into account.
\subsection{Universal optimality for parameter estimation}

Block orthogonality decouples the treatment and nuisance parameters. The least-squares estimator of \(\boldsymbol\psi\) is
\[
\hat{\boldsymbol\psi} = (\mathbf{X}^\top\mathbf{X})^{-1}\mathbf{X}^\top\mathbf{Y},
\]
and 
\[
\operatorname{Cov}(\hat{\boldsymbol\psi}) = (\mathbf{X}^\top\mathbf{X})^{-1}.
\]
Let \(\mathbf{M}_n = \mathbf{X}^\top\mathbf{X} = n\mathbf{A}_n\) with \(\operatorname{tr}(\mathbf{A}_n)=q\).  
The covariance matrix is therefore proportional to \(\mathbf{A}_n^{-1}\), and the design problem reduces to choosing \(\mathbf{A}_n\) to make this covariance as small as possible in a suitable sense.

In the realm of experimental design, a design criterion \(\Phi\) refers to a real‑valued function of the covariance matrix.  Following \cite{kiefer1974general}, \(\Phi\) satisfies
\begin{enumerate}[(i)] 
  \item~~ \textbf{convexity}. For any two positive definite matrix $\mathbf{C}_1$ and $\mathbf{C}_2$, \(\Phi(\alpha \mathbf{C}_1 + (1-\alpha)\mathbf{C}_2) \le \alpha \Phi(\mathbf{C}_1) + (1-\alpha)\Phi(\mathbf{C}_2)\) for \(\alpha\in[0,1]\);
  \item~~ \textbf{monotonicity} If  \(\mathbf{C}_1 \preceq \mathbf{C}_2\) (with respect to the Loewner order), then \(\Phi(\mathbf{C}_1) \le \Phi(\mathbf{C}_2)\);
  \item~~ \textbf{orthogonally invariant}.  \(\Phi(\mathbf{U}\mathbf{C}\mathbf{U}^\top) = \Phi(\mathbf{C})\) for every orthogonal matrix \(\mathbf{U}\).
\end{enumerate}
The corresponding spectral ordering includes the classical A-, D-, and E-optimality criteria by taking $\Phi$ as trace, determinant, and maximum eigen-value with respect to $\mathbf{A}^{-1}$. 

\begin{theorem}[Universal optimality for estimation]\label{thm:est-opt}
Assume block orthogonality, i.e.\ \(\mathbf{X}^\top\mathbf{Z}=\mathbf{0}\) and \(\mathbf{X}^\top\mathbf{J}=\mathbf{0}\).  
Among all designs with \(\mathbf{M}_n = n\mathbf{A}_n\), \(\mathbf{A}_n\succ\mathbf{0}\) and \(\operatorname{tr}(\mathbf{A}_n)=q\), the isotropic design \(\mathbf{A}_n = \mathbf{I}_q\) minimises \(\Phi\{\operatorname{Cov}(\hat{\boldsymbol\psi})\}\) for every convex, monotone and orthogonally invariant functional \(\Phi\).  
In particular, it is A‑, D‑ and E‑optimal.
\end{theorem}

Together with Theorem~\ref{thm:minimax-opt}, Theorem~\ref{thm:est-opt} establishes a \textbf{double optimality} of the block‑orthogonal isotropic design: the same allocation maximises the worst‑case growth rate of the safe test and minimises every convex orthogonally invariant measure of estimation error.  
There is no trade‑off between early stopping power and post‑experiment precision; the design actively eliminates nuisance variation and distributes information evenly, making it simultaneously optimal for both objectives laid out in Section~\ref{subsec:model}.

\subsection{Construction of the optimal design}\label{constructionoftheoptimaldesign}

The double optimality results above require a design that is both block-orthogonal and isotropic, namely,
\(\mathbf{X}^\top\mathbf{Z}=\mathbf{0}\), \(\mathbf{X}^\top\mathbf{J}=\mathbf{0}\), and \(\mathbf{X}^\top\mathbf{X}=n\mathbf{I}_q\).
We now give three explicit constructions, each tailored to a different inferential target.

All three constructions build on orthogonal arrays \citep{Hedayat1999OA}.
A two-level orthogonal array \(\mathrm{OA}(n,p,2,3)\) is an \(n\times p\) matrix with entries in \(\{-1,1\}\) such that every three-column subarray contains each of the eight possible sign patterns equally often.
Its strength-three property implies that the columns are balanced and pairwise orthogonal. Thus, for the full array \(\mathbf{A}\), \(\mathbf{A}^\top\mathbf{A}=n\mathbf{I}_p\); it also implies that the elementwise product of any three distinct columns sums to zero.

The first construction targets \(\boldsymbol\psi\) alone.
Take an \(\mathrm{OA}(n,q+1,2,3)\), use the first \(q\) columns as the treatment matrix \(\mathbf{X}\), and denote the remaining column by \(\mathbf{z}\).
For any desired number of blocking factors \(p-q\), set \(\mathbf{Z}=\mathbf{z}\mathbf{1}_{p-q}^\top\), so that every blocking column equals \(\mathbf{z}\).
Pairwise orthogonality immediately gives \(\mathbf{X}^\top\mathbf{Z}=\mathbf{0}\).
Moreover, for any treatment column \(\mathbf{x}_i\) and interaction column \(\mathbf{x}_j\odot\mathbf{z}\), simple calculation yields $\mathbf{x}_i^\top(\mathbf{x}_j\odot\mathbf{z})
=0$.
Hence \(\mathbf{X}^\top\mathbf{J}=\mathbf{0}\), and the design remains block-orthogonal and isotropic for arbitrarily large values of \(p-q\).
The trade-off is that \(\mathbf{Z}^\top\mathbf{Z}=n\mathbf{1}_{p-q}\mathbf{1}_{p-q}^\top\), so \(\operatorname{rank}(\mathbf{Z})=1\).
The construction is therefore sufficient when \(\boldsymbol\psi\) is the sole inferential target, but it does not separately identify the components of \(\boldsymbol\lambda\) or \(\boldsymbol\theta\).

To estimate the individual block main effects \(\boldsymbol\lambda\), we use a richer construction based on Walsh--Hadamard matrices.
Fix \(q\), choose powers of two \(m=2^s\ge q+1\) and \(r=2^t\), and set \(n=mr\).
Let \(\mathbf{H}_m\) and \(\mathbf{H}_r\) be Walsh--Hadamard matrices normalised so that \(\mathbf{H}_k^\top\mathbf{H}_k=k\mathbf{I}_k\).
Form \(\mathbf{X}_0\) from any \(q\) nonconstant columns of \(\mathbf{H}_m\), and let \(\mathbf{H}_r^\circ\) denote \(\mathbf{H}_r\) with its constant column removed.
Define
\(
\mathbf{X} = \mathbf{X}_0 \otimes \mathbf{1}_r,
\qquad
\mathbf{Z} = \mathbf{H}_m \otimes \mathbf{H}_r^\circ,
\)
so that \(p-q=m(r-1)=n-m\).
Kronecker-product orthogonality gives \(\mathbf{X}^\top\mathbf{X}=n\mathbf{I}_q\), \(\mathbf{Z}^\top\mathbf{Z}=n\mathbf{I}_{p-q}\), and \(\mathbf{X}^\top\mathbf{Z}=\mathbf{0}\).
The key additional fact is that Walsh--Hadamard columns are closed under elementwise multiplication.
For any treatment column \(\mathbf{x}_a\otimes\mathbf{1}_r\) and blocking column \(\mathbf{h}_k\otimes\mathbf{g}_\ell\),
\(
(\mathbf{x}_a\otimes\mathbf{1}_r)\odot(\mathbf{h}_k\otimes\mathbf{g}_\ell)
=
(\mathbf{x}_a\odot\mathbf{h}_k)\otimes\mathbf{g}_\ell,
\)
which is again a column of \(\mathbf{Z}\).
Consequently, \(\operatorname{col}(\mathbf{J})\subseteq\operatorname{col}(\mathbf{Z})\), and hence \(\mathbf{X}^\top\mathbf{J}=\mathbf{0}\).
If the model excludes treatment--block interactions, the orthogonality of \(\mathbf{Z}\) makes all \(p-q\) block main effects separately estimable.
If interactions are included, \(\boldsymbol\psi\) remains optimally estimable, but \(\boldsymbol\lambda\) and \(\boldsymbol\theta\) are aliased because \(\operatorname{col}(\mathbf{J})\subseteq\operatorname{col}(\mathbf{Z})\).
Since \(p-q=n-m\), this construction accommodates a number of blocking factors close to the sample size.

The third construction is appropriate when the interaction parameters \(\boldsymbol\theta\) must also be identified.
Neither preceding construction suffices for this purpose: the repeated-column design does not identify the block effects separately, whereas the Walsh--Hadamard design aliases \(\boldsymbol\lambda\) with \(\boldsymbol\theta\).
Instead, take a full strength-three array \(\mathrm{OA}(n,p,2,3)\), using its first \(q\) columns for \(\mathbf{X}\) and the remaining \(p-q\) columns for \(\mathbf{Z}\).
Then \(\mathbf{X}^\top\mathbf{X}=n\mathbf{I}_q\), \(\mathbf{Z}^\top\mathbf{Z}=n\mathbf{I}_{p-q}\), and \(\mathbf{X}^\top\mathbf{Z}=\mathbf{0}\).
Every interaction column \(\mathbf{x}_j\odot\mathbf{z}_k\) is also orthogonal to every treatment column \(\mathbf{x}_i\): when \(i\neq j\), this follows from strength three, and when \(i=j\), it follows from the balance of \(\mathbf{z}_k\).
Thus \(\mathbf{X}^\top\mathbf{J}=\mathbf{0}\).
Unlike in the Walsh--Hadamard construction, the interaction columns need not lie in \(\operatorname{col}(\mathbf{Z})\); therefore, \(\boldsymbol\lambda\) and \(\boldsymbol\theta\) are separately identifiable whenever the combined design matrix \([\mathbf{X},\mathbf{Z},\mathbf{J}]\) has full column rank.
This full-rank requirement entails \(n\ge p+q(p-q)\), which is of order \(p^2\) when both \(q\) and \(p-q\) grow proportionally with \(p\).

The first two constructions are especially useful when the nuisance dimension is comparable to the sample size.
The repeated-column design permits arbitrarily many blocking columns, although their effects are not separately identifiable; the Walsh--Hadamard construction satisfies \(p-q=n-m\), so \(p/n\to1\) for fixed \(m\) as \(n\) grows.
In both cases, explicit combinatorial orthogonality preserves the optimal e-value growth rate \(B_n(\rho_n)\), regardless of the nuisance dimension.
This contrasts sharply with the uncontrolled i.i.d.\ design considered in safe testing proposed in \cite{grunwald2020safe} (see Proposition~\ref{prop:fail}), for which nuisance factors cause a first-order loss in the minimax rate.
The comparison illustrates the principle of central design-driven: when \(\boldsymbol\psi\) is the inferential target, nuisance variation should be made orthogonal to the treatment columns by design, rather than handled through post hoc conditioning or marginalization.
The full strength-three construction extends this protection to the separate estimation of \(\boldsymbol\lambda\) and \(\boldsymbol\theta\), at the cost of a larger sample size.

\vspace{-1em}
\section{Sequential Implementation and Anytime-Validity}
\label{subsec:sequential}

Section~\ref{sec:joint} addresses the joint optimality of testing and
estimation in a single-batch design. We now return to Objectives~I and~II
in Section~\ref{subsec:model} and extend the analysis to a sequential
framework in which the experiment consists of batches \(t=1,2,\ldots\),
each of size \(n_t\), and the experimenter may terminate the study early
by monitoring the experimental output after each batch. Under the null,
for every \(\boldsymbol\psi\) with \(\|\boldsymbol\psi\|\ge\Delta\), any
e-value sequence \(\{E_t\}\) satisfying
\(\mathbb{E}_{\boldsymbol\psi}[E_t\mid\mathcal{F}_{t-1}]\le 1\) for all
\(t\ge1\) ensures that the running product
\(E^{(T)}=\prod_{t=1}^{T}E_t\) is a nonnegative test supermartingale.
By Ville's inequality, for any stopping time \(\tau\) adapted to the
underlying filtration,
\(\mathbb{P}(E^{(\tau)}\ge 1/\alpha)\le \alpha\), so the type I error
rate is controlled under optional stopping. A natural choice for \(E_t\)
is the exact radial e-value \(E_{n_t}^*\) from
Theorem~\ref{thm:optimal-e}, applied to the score from batch \(t\).

For Objective~I, the sequential problem is to accumulate evidence rapidly
when the alternative \(H_1\) holds, while retaining time-uniform type I
error control. Here the e-value betting interpretation is useful
\citep{Shafer2021testing}. To compare different sequential betting
strategies, however, we need a criterion for the speed of evidence
accumulation. Kelly's expected log-growth criterion \citep{kelly1956} is
the natural benchmark: maximizing conditional expected log-growth is
equivalent to maximizing the exponential rate at which evidence is
gathered.

Let \(\pi\in\Pi\) be a strategy that, before batch \(t\), selects a design
\(D_t(h)\in\mathfrak D_{n_t}^{\perp}\) and a nonnegative score factor
\(e_t(h,\cdot)\) as jointly measurable, predictable functions of the
observed history \(h=\mathbf H_{t-1}\). In line with
Section~\ref{subsec:e-value-comp}, the design class
\(\mathfrak D_{n_t}^{\perp}\) contains the feasible block-orthogonal
designs \(D\) with
\(\mathbf A(D)=\mathbf X_D^\top\mathbf X_D/n_t\succ0\) and
\(\operatorname{tr}\mathbf A(D)=q\). We measure the sequential evidence
accumulation of \(\pi\) by the following criterion.

\begin{definition}[Cumulative conditional worst-case growth]
\label{def:sequential-grow}
For a joint strategy \(\pi\in\Pi\), define the cumulative conditional
worst-case growth by
\(\mathcal C_T^\pi(\rho):=\sum_{t=1}^T c_t^\pi(\mathbf H_{t-1};\rho)\),
with
\(c_t^\pi(h;\rho):=\inf_{\|\boldsymbol\vartheta\|\le\Delta-\rho}
\mathbb E_{\boldsymbol\vartheta}^{D_t(h)}[\log e_t(h,\mathbf S_t)]\).
\end{definition}

The following proposition states that the batchwise greedy rule is not
merely a heuristic but also sequentially optimal for cumulative
conditional worst-case log-growth. Recall that
\(V_T(\rho):=\sum_{t=1}^T
\sup_{D\in\mathfrak D_{n_t}^{\perp}}
\Gamma_{n_t}(\mathbf A(D),\rho)\).

\begin{proposition}[Batchwise greedy optimality]
\label{prop:batchwise-greedy}
Assume that, for each \(t\) and \(h\), the single-batch supremum
\(\sup_{D\in\mathfrak D_{n_t}^{\perp}}\Gamma_{n_t}(\mathbf A(D),\rho)\)
is either attained by an admissible pair \((D_t^\star(h),e_t^\star(h,\cdot))\)
or can be approximated measurably up to any \(\varepsilon>0\). Then,
pathwise,
\(\sup_{\pi\in\Pi}\mathcal C_T^\pi(\rho)=V_T(\rho)\).
Moreover, the batchwise greedy strategy that at every history chooses a
single-batch optimal, or \(\varepsilon\)-optimal, pair satisfies
\(\mathcal C_T^{\pi^\star}(\rho)=V_T(\rho)\) or
\(\mathcal C_T^{\pi^\star}(\rho)\ge V_T(\rho)-\varepsilon\),
respectively.
\end{proposition}

This conclusion follows from the additive structure of the objective. For
a strategy \(\pi\in\Pi\), define the running product
\(E^{\pi,(T)}=\prod_{t=1}^T e_t(\mathbf H_{t-1},\mathbf S_t)\), with
\(E^{\pi,(0)}=1\), where \(e_t\) is the score factor selected by \(\pi\).
Then
\(\log E^{\pi,(T)}=\sum_{t=1}^T\log e_t(\mathbf H_{t-1},\mathbf S_t)\).
Because the validity constraint is imposed conditionally batch by batch,
and \(n_t\) and \(\mathfrak D_{n_t}^{\perp}\) are exogenous and batchwise
separable, the Bellman recursion for
\(\sup_{\pi\in\Pi}\mathcal C_T^\pi(\rho)\) has no continuation value: the
future value function does not depend on the current choice of
\((D_t,e_t)\). Consequently, the optimal action is to maximize the current
batch's conditional worst-case log-growth. This is the conditional
worst-case analogue of Kelly's myopic optimality.

Proposition~\ref{prop:batchwise-greedy} yields a simple design and
analysis principle: the sequential problem admits a myopic solution. It
suffices to choose, in each batch, a single-batch optimal
design--e-value pair and then multiply the resulting e-values. This
motivates a simple implementation in which the same block-orthogonal
isotropic design is repeated independently across batches and the
batchwise e-values are multiplied.

When all batch sizes are equal, say \(n_t\equiv n\), the above
prescription becomes a pure repetition strategy: repeat the same
single-batch design--e-value rule independently across batches and multiply
the resulting e-values. Concretely, fix a separation \(\rho_n\) in advance
and put \(\delta_0=n\Delta^2\) and
\(\delta_1=n(\Delta-\rho_n)^2\). In every batch, use the same
block-orthogonal isotropic design and the same exact conditional e-value
rule. With \(Q_t=n^{-1}\|\mathbf S_t\|^2\), the batchwise factor from
Theorem~\ref{thm:optimal-e}, evaluated using the score from batch \(t\), is
\(E_{n_t}^*=f_{q,\delta_1}(Q_t)/f_{q,\delta_0}(Q_t)\). To distinguish this
single-batch factor from the cumulative e-process, define the sequential
evidence by
\(\mathcal E_T^*=\prod_{t=1}^T E_{n_t}^*\), with
\(\mathcal E_0^*=1\). Write
\(B_n^{\mathrm{joint}}(\rho):=\sup_{D\in\mathfrak D_n^{\perp}}
\Gamma_n(\mathbf A(D),\rho)\). Since \(B_n^{\mathrm{joint}}(\rho)\) is the
single-batch supremum, the cumulative benchmark reduces to
\(V_T(\rho)=T B_n^{\mathrm{joint}}(\rho)\).

The replicative construction also preserves the estimation optimality of
Objective~II at the terminal horizon. Because the same block-orthogonal
isotropic design is used in every batch, repetition does not alter the
orthogonality conditions: the cross terms with the nuisance regressors
vanish batch by batch, and the total treatment information is
\(\sum_{t=1}^T \mathbf X_t^\top \mathbf X_t
= \sum_{t=1}^T n_t \mathbf I_q
= N_T \mathbf I_q\), where \(N_T=\sum_{t=1}^T n_t\). Consequently, the
terminal estimator based on the complete experiment,
\(\widehat{\boldsymbol\psi}_T
= N_T^{-1}\sum_{t=1}^T \mathbf S_t\), satisfies
\(\widehat{\boldsymbol\psi}_T
\sim N(\boldsymbol\psi, N_T^{-1}\mathbf I_q)\). Thus the universal
optimality of the single-batch design, including A-, D-, and
E-optimality, is inherited at the terminal horizon among prespecified
block-orthogonal schedules with total information trace \(N_Tq\). In this
sense, repetition preserves orthogonality and carries the estimation
optimality of Objective~II forward to the full experiment without any
additional adaptive redesign.

The optimality in both testing and estimation is summarized in the
following theorem.

\begin{theorem}[Multi-batch double optimality]
\label{thm:sequential-grow}
Put \(N_T=\sum_{t=1}^T n_t\) and let
\(\mathcal C_T^*\) denote the growth guarantee of the strategy
based on the exact radial factor. At every deterministic horizon \(T\):
\begin{enumerate}[(i)]
\item \textbf{Sequential testing.}
For \(n_t\equiv n\) with \(n\to\infty\) and \(0<\rho_n\to0\),
\(\mathcal C_T^*(\rho_n)/V_T(\rho_n)\to1\), uniformly over
\(T\ge1\). The exact-factor strategy is therefore first-order
sequentially optimal.
\item \textbf{Terminal estimation.} In the prespecified complete
experiment, the proposed design yields
\(\widehat{\boldsymbol\psi}_T=N_T^{-1}\sum_{t=1}^T\mathbf S_t
\sim N(\boldsymbol\psi,N_T^{-1}\mathbf I_q)\).
It is universally optimal in the sense of Theorem~\ref{thm:est-opt}
among prespecified block-orthogonal schedules with positive definite total
treatment information of trace \(N_Tq\).
\end{enumerate}
\end{theorem}

In practice, the repeated structure greatly simplifies implementation. The
same batchwise design is reused at every stage, so the full allocation
schedule can be precomputed in advance, eliminating online optimization
and reducing operational overhead. The batchwise design itself can be
constructed from orthogonal arrays of strength at least three, as
described in Section~\ref{subsec:oa}; such arrays are widely available in
run sizes considerably smaller than a full factorial. For example, an
experiment with seven factors (\(p=7\)) requires \(128\) runs per batch
under a full factorial, whereas an orthogonal array of strength three can
retain the required orthogonality and estimation properties with \(16\)
runs. This fixed, precomputed schedule with small, flexible batch sizes
makes the approach readily deployable in large-scale online experiments.

{
The preceding results establish that the repeated block-orthogonal
isotropic design is anytime-valid and attains the sequential benchmark.
The following proposition explains why predictable adaptation cannot
improve on this choice: in the critical regime, the single-batch optimum
is governed by the least favourable local information, which adaptation
can redistribute across directions but not increase.
\begin{proposition}
[No adaptive improvement in local minimax information]
\label{thm:adapt-no-gain}
Among predictable block-orthogonal designs with the same batch budget, the
fixed isotropic design maximizes, conditionally on every history, the
least favourable critical information for \(H_{1,n_t}(\rho_{n_t})\).
Equipped with the exact e-value \(E_{n_t}^*\), it attains the local GROW upper
bound \(h^2/2+o(1)\) when \(\sqrt{n_t}\rho_{n_t}\to h\).
\end{proposition}}

{A more aggressive comparator replaces radial evidence by a predictable
data-directed linear likelihood ratio. Such a construction is not uniformly
valid for the exterior null. For any predictable nonzero betting vector
\(\mathbf w_t\), conditional validity would require
\(\mathbf w_t^\top\mathbf M_t\boldsymbol\psi\le0\) for every
\(\|\boldsymbol\psi\|\ge\Delta\). This is impossible because the null
contains both \(\boldsymbol\psi\) and \(-\boldsymbol\psi\).
Predictability prevents reuse of the current response, but it cannot turn
a directionally calibrated linear likelihood ratio into an e-value for the
two-sided exterior null. We therefore use this construction only as an
invalidity diagnostic in the numerical comparison, not as a separate
inferential procedure.}

{The analyses above deliver a sharp message. Adaptive design can provide
pointwise gains in a selected direction, but it cannot improve the
direction-free local minimax growth rate and may weaken estimation in
non-favoured directions. Data-directed linear betting raises a different
issue: it is not an e-value uniformly over the exterior null. The fixed
isotropic allocation combined with radial evidence therefore
retains validity, attains the critical minimax growth bound, and preserves
universal estimation optimality simultaneously.}

	\section{Numerical Studies}
	\label{sec:numerical}

    We consider the linear model
\begin{equation}\label{eq:model-sim1}
Y_i = \mathbf{x}_i^{\top} \boldsymbol\psi + \mathbf{z}_i^{\top} \boldsymbol\lambda + (\mathbf{x}_i \otimes \mathbf{z}_i)^{\top} \boldsymbol\theta + \varepsilon_i,
\quad \varepsilon_i \sim N(0,1),
\end{equation}
where $\mathbf{x}_i \in [-1,1]^2$ is the treatment vector, 
$\mathbf{z}_i = (Z_{i1}, Z_{i2}, Z_{i3}, Z_{i4})^{\top}$ contains two discrete and two continuous nuisance factors, which lies in $\{-1,+1\}$ and $[-1,1]$, respectively.
The nuisance and interaction coefficients are chosen with deliberately unequal magnitudes to mirror the varying importance of contextual factors in practice:
\[
\begin{aligned}
\boldsymbol\lambda &= (0.80,\,-0.60,\,0.45,\,-0.35)^{\top},\\
\boldsymbol\theta &= (0.55,\,-0.45,\,0.40,\,-0.35,\,-0.30,\,0.28,\,-0.25,\,0.22)^{\top}.
\end{aligned}
\]
Several entries of $\boldsymbol\lambda$ and $\boldsymbol\theta$ are substantially larger than the treatment effects we aim to detect (specified below), so the signal of interest is embedded in strong nuisance-driven variation---a challenging but realistic feature.
 The testing problem is that 
\(
\mathcal{H}_0: \|\boldsymbol\psi\| \ge \Delta
\quad\text{against}\quad
\mathcal{H}_1: \|\boldsymbol\psi\|<\Delta,
\)
with practical threshold $\Delta = 0.65$.

Throughout, we report the following metrics.
Under null hypothesis ($\mathcal{H}_0$):
\begin{itemize}
    \item[-] \textbf{False stopping rate (FSR)}: the probability that the 
    experiment stops before all $T$ batches are collected when the null 
    is true.
    \item[-] \textbf{Overall type-I error (Type-I)}: the probability of falsely rejecting the null. It consists of false rejections that occur at 
    an early stopping time and false rejections after spending out all the design budgets. This is controlled directly by the proposed e-value and FSR is under control when we rein this type-I error. 
    \item[-] \textbf{Mean squared error (MSE)}: under the simple null we 
    report the MSE of the block effect $\bm\lambda$; in the exploratory 
    experiment we additionally report the MSE of the treatment effect 
    $\bm\psi$, computed under the composite null with the full horizon.
\end{itemize}
Under the alternative hypothesis ($\mathcal{H}_1$):
\begin{itemize}
    \item[-] \textbf{Average stopping batch (Stop)}: the expected number of batches 
    at which the experiment terminates when the alternative is true. 
    Smaller values indicate faster release of experimental capacity.
    \item[-] \textbf{Expected log-growth (Growth)}: the theoretical one-batch expected log-growth 
    of the e-value under the alternative. This serves as a measure of 
    test power within the e-value framework.
\end{itemize}

	We compared the proposed design presented in Section~\ref{sec:joint}, with following competitors.
    \begin{enumerate}[(I)]
        


        

           \item\quad  Each element of
	        $(x_1,x_2,z_3,z_4)$ is drawn independently from
	        \(\operatorname{Unif}[-1,1]\), while \(z_1,z_2\) are independent
	        Rademacher variables. Conditional on each realized batch design, we
	        project $\mathbf X_t$ and $\mathbf Y_t$ off the nuisance design
	        $[\mathbf Z_t,\mathbf J_t]$ and apply the safe
	        testing procedure in \cite{grunwald2020safe}. The resulting batchwise
	        factors are multiplied across batches.

	        	\item\quad Adaptive design. The optimal design under $\mathcal{H}_0:\bm\psi=\bm\delta$ against $\mathcal{H}_1:\bm\psi=\bm 0$, with $\bm\delta$ chosen as $\hat{\boldsymbol\psi}_{t-1}$. For the first batch, before an OLS estimate is available, we initialize the direction randomly. The inference procedure uses the least-information exact radial factor defined above.
	\item\quad Adaptive linear design. The design is the same as (IV), while the inference procedure corresponds to the adaptive betting with the linear e‑value. The first-batch betting vector is drawn from the prior distribution $\hat{\boldsymbol\psi}_0\sim N\!\left(\boldsymbol\psi,q^{-1}\mathbf I_q\right)$.

    \end{enumerate}

All designs share the same per-batch sample size with batch size $n=32$, batch number $T=8$, and monitoring level with $\alpha=0.05$. 
We evaluate the performance base on the above metrics empirically via 1,000 independent replications.

\begin{table}[htbp]
	\centering
	\caption{\textbf{Exploratory composite-null experiment.}}
	\label{tab:composite-sim}
	{
		\begin{tabular}{lcccc@{\hspace{0.8em}}c@{\hspace{0.8em}}cc}
			\hline
			& \multicolumn{4}{c}{$\mathcal{H}_0$} & & \multicolumn{2}{c}{$\mathcal{H}_1$}\\
			\cline{2-5}\cline{7-8}
			Procedure & FSR & Type-I & MSE$_\psi$ & MSE$_\lambda$ & & Growth & Stop\\
			\hline
Proposed & 1.4 & 5.8 & 0.0080 & 0.0163 & & 3.704 & 1.406 \\
(I) & 0.7 & 3.8 & 0.0257 & 0.0341 & & 0.165 & 7.733 \\
(II) & 0.1 & 0.2 & 0.0197 & 0.9654 & & 0.554 & 5.448 \\
(III) & 19.6 & 19.6 & 0.0195 & 0.9725 & & NA & 7.979 \\
\hline
	\end{tabular}}
\end{table}

Table~\ref{tab:composite-sim} gives the composite-null comparison.The Growth entry for adaptive linear design is reported as NA because it uses a data-directed linear betting e-value and thus the results are not comparable.
Relative to procedure~I proposed in \cite{grunwald2020safe}, our proposed method increases expected log-growth and reduces the average stopping batch. Its treatment- and nuisance-effect MSEs also decrease. The proposed design enforces
block orthogonality and isotropy, whereas complete randomization in (I) loses
treatment information through chance alignment with the nuisance regressors
before the projection step. 
Procedures~II and~III illustrate the two cautions surrounding
Proposition~\ref{thm:adapt-no-gain}. Procedure~II predictably redirects
information toward the estimated direction and retains a valid radial factor,
but performs worse than Proposed in growth, stopping, and treatment-effect
estimation. This pattern is qualitatively consistent with the proposition that
directional adaptation cannot improve least-favourable local information,
although the finite-horizon simulation is not a direct verification of that
local asymptotic result. Procedure~III uses the same adaptive design with a
data-directed linear bet and exhibits substantial error inflation, illustrating
the separate warning that predictability alone does not ensure uniform validity
over the two-sided exterior null.


  \section{Real Data Analysis}
\label{sec:real-data}

We now apply the proposed framework to a prompt-design experiment for a large language model.
The managerial problem is increasingly common: as inference tokens become costly, a platform wants to identify effective prompt templates quickly rather than iterating through long prompt variants by trial and error.
We treat prompt design as a multifactor online experiment and use answers produced by DeepSeek's chat-completion API using the \texttt{deepseek-v4pro} model to compare alternative experimental designs under the same response budget.
\subsection{DeepSeek Prompt Experiment}
\label{subsec:deepseek-data}
The task is a proof-based mathematical statistics exercise:
let \(X_1,\ldots,X_n\) be iid \(N(\mu,\sigma^2)\), where \(\sigma^2\) is known, and prove using the Cramer--Rao lower bound that \(\bar X\) is the minimum-variance unbiased estimator of \(\mu\).
This exercise is useful for prompt evaluation because a correct answer must provide a chain of proof steps rather than only a numerical final answer.
The four factors are assigned as follows:
\[
\begin{aligned}
X_1 &= +1 \text{ (role after the problem)}, &\quad\text{or}\quad& -1 \text{ (role before)};\\
X_2 &= +1 \text{ (includes mathematical formulas)}, &\quad\text{or}\quad& -1 \text{ (no formulas)};\\
Z_1 &= +1 \text{ (long prompt)}, &\quad\text{or}\quad& -1 \text{ (short prompt)};\\
Z_2 &= +1 \text{ (precise role description)}, &\quad\text{or}\quad& -1 \text{ (vague role description)}.
\end{aligned}
\]

The \(Z\)-factors are not the main managerial levers in this application; they capture prompt verbosity and the specificity of the persona, both of which can affect answer quality and token cost and therefore should not be allowed to confound the two role-design effects.
The primary estimands are the two treatment contrasts associated with \(X\): whether moving the role statement changes performance and whether including mathematical formulas improves performance.
The \(Z\)-contrasts are treated as block or context effects: they are estimated to diagnose how verbosity and role precision shift score and cost, but they are not the main prompt-design decision.
We analyze the response pool under the composite-null testing regime studied throughout the paper.

Each prompt was sent to DeepSeek in 8 complete batches, so every prompt cell has eight independent responses at temperature \(0.6\).
Scores were assigned by a deterministic rule-based rubric specified before analysis rather than by ex post subjective judgment.
The rubric gives one point for each of six proof components: unbiasedness of \(\bar X\), variance of \(\bar X\), the score function, Fisher information, the Cramer--Rao lower bound, and the conclusion that \(\bar X\) attains the bound and is therefore the minimum-variance unbiased (MVU) estimator.
The experiment also records total tokens and response latency.
All design comparisons use the token-adjusted utility
\(
U=\text{proof score}-\text{total tokens}/1000.
\)
After each complete batch, we compute the treatment-score contrast implied by the sampled design and apply the composite-null exact e-value.
We model the token-adjusted utility as
\(
U_i = \mathbf{x}_i^{\top}\boldsymbol\psi + \mathbf{z}_i^{\top}\boldsymbol\lambda + (\mathbf{x}_i\otimes\mathbf{z}_i)^{\top}\boldsymbol\theta + \varepsilon_i,
\qquad \varepsilon_i \sim N(0,\sigma^2),
\)
where $\mathbf{x}_i, \mathbf{z}_i$ are the $\pm 1$ treatment and nuisance vectors for the $i$th prompt.

In the real-data experiment the true treatment effect is unknown, so we formulate the practical composite-null hypothesis
\(
\mathcal{H}_0:\|\boldsymbol\psi\|\ge 0.5
\qquad\text{versus}\qquad
\mathcal{H}_1:\|\boldsymbol\psi\|<0.5.
\)
The threshold \(0.5\) corresponds to roughly half a rubric point on the token-adjusted utility scale and represents the smallest effect a manager would consider practically meaningful.
The reported sequential exact e-value uses the centre calibration \(\rho_n=0.5\); it remains valid for the exterior null, while its growth target is the centre of the composite alternative.

\subsection{Results}
\label{subsec:deepseek-results}

The experiment uses the same 128 responses (16 prompt cells, 8 batches) and the same design rules as in the simulation.
For each design we record the stopping batch, the log-e-value at stopping or at the full horizon, and the statistical decision at \(\alpha=0.05\); experiments that do not stop are marked as ``Full horizon.''
The results are collected in Table~\ref{tab:prompt-design-comparison}.

\begin{table}[H]
	\centering
	\caption{\textbf{Real-data procedure comparison under the composite null hypothesis.}}
	\label{tab:prompt-design-comparison}
	{
		\begin{tabular}{llll}
			\hline
			Procedure & Stop batch & Decision & Log e-value\\
			\hline
				Proposed & 4 & Reject & 3.14 \\
(I) & Full horizon & Accept & -14.24 \\
(II) & Full horizon & Accept & 1.64 \\
(III) & Full horizon & Accept & -11.43 \\
\hline
		\end{tabular}}
\end{table}

Under the composite null, only our design rejects \(\mathcal{H}_0\), whereas the remaining designs fail to reject and exhaust the full budget.
This divergence is consistent with the actual signal strength: the full-data benchmark on all 128 responses (Table~\ref{tab:prompt-full-data-test}) rejects the practical-effect boundary, and the estimated norm \(\|\hat{\boldsymbol\psi}\|=0.288\) falls well below the managerially meaningful threshold \(\Delta_\psi=0.50\).
The signal is simply too weak for low-powered designs to resolve within the budget; our design, which concentrates information more efficiently, reaches a valid rejection, while the competing designs exhaust all batches without a conclusion.

\begin{table}[htbp]
	\centering
	\caption{\textbf{Full-data composite-null benchmark using all DeepSeek responses.}}
	\label{tab:prompt-full-data-test}
	{\small
		\begin{tabular}{llrrrrrl}
			\hline
			Test & Null & \(\hat\psi_1\) & \(\hat\psi_2\) & \(\|\hat{\boldsymbol\psi}\|\) & \(F\)-Statistic & \(p\)-value & Decision\\
			\hline
	Practical-effect boundary & $\mathcal{H}_0:\|\boldsymbol\psi\|\ge \Delta_\psi$ & -0.055 & 0.283 & 0.288 & 4.97 & 0.014 & Reject \\
\hline
	\end{tabular}}
\end{table}

This outcome is not merely a methodological demonstration, it carries a substantive finding.
Role placement and formula inclusion, despite their prominence in prompt-engineering practice, produce only a modest utility gain, likely too small to justify dedicated optimization effort in comparable reasoning tasks.
The proposed design delivers this insight early and at lower cost; the competing designs spend the full budget searching for a signal that, as the data reveal, is simply not large enough to matter.
In such weak-signal regimes, which are arguably the norm in mature online experimentation, the choice of design directly determines whether the team obtains a timely answer or exhausts resources without one.

\section{Conclusion}
This paper develops an unified, design-driven inference framework for multifactor online experimentation with optional stopping. The central insight is that validity under continuous monitoring is not merely a property of the stopping rule. It is fundamentally shaped by the information geometry created before the first observation is collected. By combining block-orthogonal designs with e-value stopping rules, our framework achieves error control while eliminating nuisance block effects by construction. Crucially, the orthogonal structure absorbs or cancels out interaction effects, so that main-effect inference remains valid even in the presence of non-zero interactions. This design-driven strategy inverts the usual logic: instead of correcting for confounding after data are generated, we tailor the data-collection process so that sequential inference operates at full efficiency without post-hoc adjustment.

Our analysis identifies two distinct managerial regimes, each requiring a fundamentally different experimental design. When the anticipated effect direction is known (confirmatory testing), the optimal design concentrates information along that specific direction. In this regime, replicating the same treatment allocation is a source of speed and statistical power rather than waste, allowing the e-value to accumulate evidence rapidly. Conversely, when the direction is unknown and the objective is robust screening over the practical-equivalence ball, we prove that the minimax optimal design is isotropic. In this regime, adaptive redirection cannot improve the least favourable boundary-local information and weakens protection in another direction. This contrast yields a clear and operational design prescription: platforms do not need complex adaptive traffic-allocation engines to support both objectives. They only need to pre-configure a block-orthogonal template and switch between directed and isotropic allocations based on the managerial objective.

The proposed framework allows for immediate implementation using established factorial or orthogonal-array constructions. Under this ANOVA-based specification, optimal allocations are available in closed form, avoiding the computational complexity of iterative numerical optimization. Two premises define the current scope: Gaussian errors are assumed, and the block-orthogonal design, while protecting main-effect tests from interactions, does not identify the interactions themselves—a trade-off that is natural when the managerial focus is on robust screening of main effects. Natural directions for future work include extending this design-driven framework to generalized linear models for non-Gaussian outcomes, incorporating designs that preserve main-effect robustness while allowing estimation of pre-specified interactions, and verifying its efficacy in prospective, large-scale deployments.

\appendix
\renewcommand{\theproposition}{A.\arabic{proposition}}
\setcounter{proposition}{0}

\section*{Appendix: Brief introduction of orthogonal array}\label{subsec:oa}

As the orthogonal array is a key ingredient in constructing the optimal design, we briefly introduce it here.
Following \cite{Hedayat1999OA}, an $N\times m$ matrix with entries from $\{-1,1\}$ is called a two-level orthogonal array of strength $t\geq 1$, denoted by $\text{OA}(N, m, 2, t)$, if every $N\times t$ sub-matrix contains each of the $2^t$ possible ordered $t$-tuples of $\pm 1$ exactly $N/2^t$ times.
A $2^m$ full factorial design is an $\text{OA}(2^m, m, 2, m)$, and more generally, any $\text{OA}(N, m, 2, t)$ with $t\geq 2$ is automatically an orthogonal array of strength two.
In practice, the strength parameter is typically chosen as $t=2$, which yields $\mathbf{A}^\top\mathbf{A}=N\mathbf{I}_m$ while keeping the number of runs $N$ substantially smaller than $2^m$.
Unless otherwise noted, we take $t=2$ throughout the paper.
When the required batch size $n_t$ exceeds the number of rows $N$ of a standard orthogonal array, one can simply replicate the array by stacking copies of it vertically; the resulting $n_t\times m$ matrix remains an orthogonal array of the same strength.

\section*{Competing interests}
The authors declare no competing interests.

\section*{Author contributions statement}
All authors contributed equally to this work and are listed alphabetically by surname.

\bibliographystyle{abbrvnat}
\bibliography{report}

\end{document}